\documentclass[sigplan,screen,10pt, nonacm]{acmart}

\renewcommand\footnotetextcopyrightpermission[1]{}
\setcopyright{none}
\AtBeginDocument{
  }

\usepackage{algorithmic}
\usepackage{graphicx}
\usepackage{textcomp}
\usepackage{array}
\usepackage{xcolor}
\usepackage{subcaption}
\usepackage{booktabs}
\usepackage{multirow}
\usepackage{listings}
\usepackage{adjustbox}
\usepackage{tabularx}
\usepackage{booktabs}
\usepackage[ruled,vlined]{algorithm2e}
\usepackage{hyperref}
\usepackage{cleveref}
\usepackage{balance}
\usepackage{placeins}

\newcommand{\ignore}[1]{}

\newcommand{\name}{SIMTurn}
\DeclareRobustCommand{\noyear}[3]{}

\newcolumntype{L}[1]{>{\raggedright\arraybackslash}p{#1}}

\begin{document}

\title{\name{}: Lightweight Divergent Subgroup Scheduling for Intra-Warp Synchronization}

\author{Nikhil Rout}
\authornote{Equal contribution}
\affiliation{%
  \institution{University of California, Los Angeles}
  \city{Los Angeles}
  \state{CA}
  \country{USA}
}
\email{nikhilrout97@ucla.edu}

\author{Joe Samuel Rosh}
\authornotemark[1]
\affiliation{%
  \institution{University of California, Merced}
  \city{Merced}
  \state{CA}
  \country{USA}
}
\email{jrosh@ucmerced.edu}

\author{Blaise Tine}
\affiliation{%
  \institution{University of California, Los Angeles}
  \city{Los Angeles}
  \state{CA}
  \country{USA}
}
\email{blaisetine@cs.ucla.edu}

\crefformat{section}{\S#2#1#3}
\crefname{section}{Section}{Sections}
\crefname{figure}{Figure}{Figures}
\crefname{table}{Table}{Tables}
\crefname{alg}{Algorithm}{Algorithms}
\crefname{equation}{Equation}{Equations}
\creflabelformat{equation}{#2\textup{#1}#3}
\crefname{appendix}{Appendix}{Appendices}
\crefname{algorithm}{Algorithm}{Algorithms}
\crefname{lstnumber}{line}{lines}
\crefname{lstnumber}{Line}{Lines}
\crefname{lstlisting}{listing}{listings}
\Crefname{lstlisting}{Listing}{Listings}
\crefname{listing}{Listing}{Listings}
\Crefname{listing}{Listing}{Listings}
\renewcommand{\crefrangeconjunction}{--}

\begin{abstract}
Stack-based reconvergence enables efficient SIMT execution with a
shared warp program counter and compact control-flow state. However,
its fixed serialization of divergent paths can deadlock programs that
rely on intra-warp synchronization: a spinning subgroup may prevent
a resource-holding sibling from executing. Independent Thread
Scheduling (ITS) enables finer-grained progress through per-thread
execution state, but introduces additional hardware cost.

We present SIMTurn, a lightweight compiler--hardware mechanism that
enables forward progress among divergent subgroups while retaining
stack-based reconvergence. SIMTurn combines a compiler-inserted yield
instruction at potentially blocking loops with a bounded pool of
resumable subgroups. The scheduler allows a waiting subgroup
to yield to a sibling that can make progress, without requiring
independent program counters for every thread.

We implement SIMTurn in the compiler, simulator, and RTL on the
open-source Vortex GPGPU and evaluate it against baseline IPDOM and per-thread ITS
schemes. Across the
evaluated GPU thread configurations, SIMTurn introduces only 17--24\%
scheduler-area overhead relative to IPDOM, compared with 39--91\%
for ITS, and stays within 8\% of IPDOM on workloads that never
need cross-subgroup progress. On kernels with genuine cross-subgroup
parallelism, SIMTurn tracks ITS closely and is up to 2.7$\times$ faster than a
software-serialized IPDOM baseline. 
\end{abstract}



\maketitle

\section{Introduction}

\begin{figure}
    \centering
    \includegraphics[width=\linewidth]{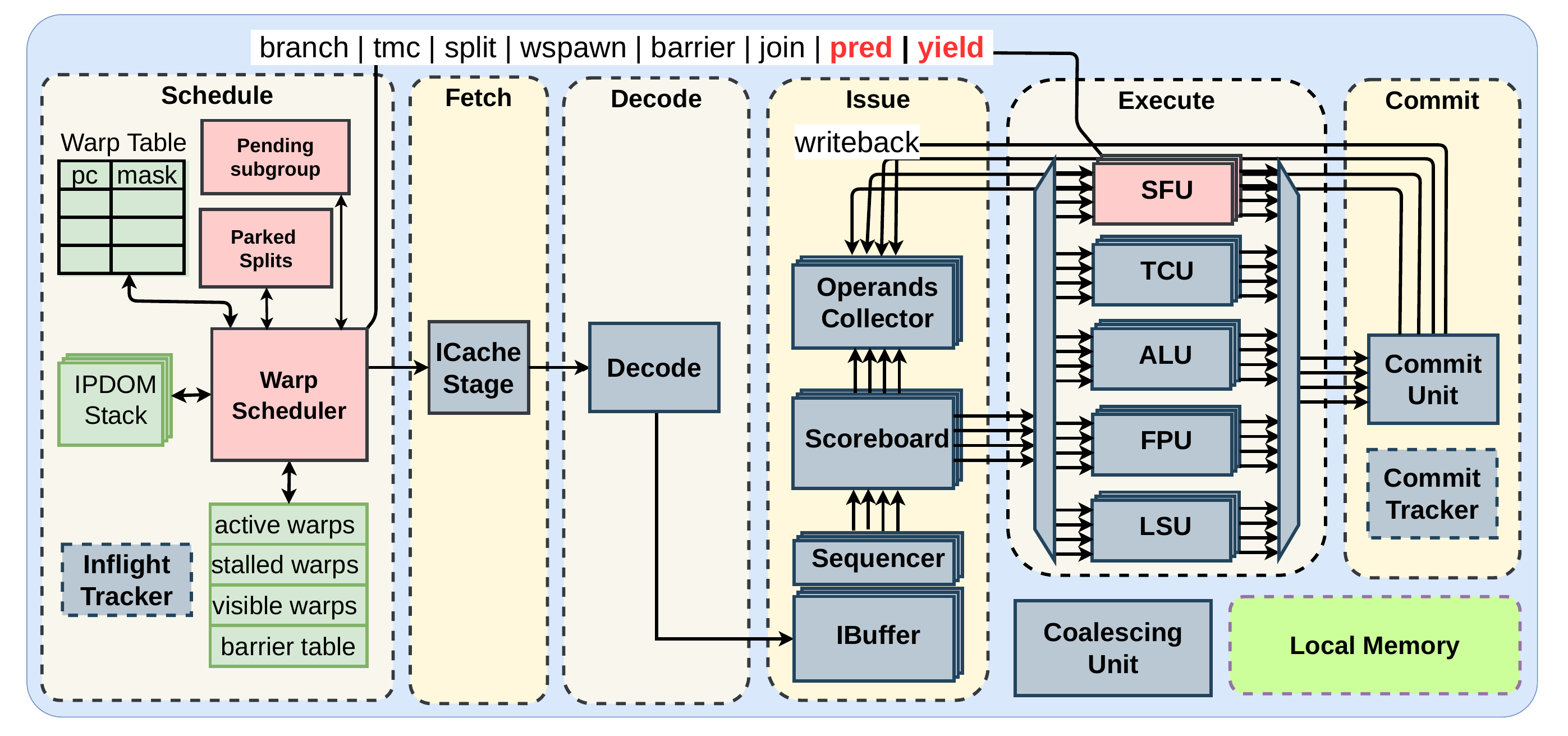}
    \caption{\name{} ISA and Microarchitecture modifications to a GPU core pipeline.}
    \label{fig:SCS_TOP}
\end{figure}

GPUs execute scalar threads under the single-instruction, multiple-thread
(SIMT) model. Threads are grouped into warps that share one instruction
stream, so control state is kept once per warp instead of once per thread.
When threads in a warp take different branch directions, the warp diverges
into subgroups. Stack-based reconvergence runs one path at a time and
merges the subgroups at the immediate post-dominator (IPDOM) of the
branch~\cite{habermaier2012simt,eltantawy2016mimd}. It needs only a shared
program counter, an active-thread mask, and a compact stack per warp.

\begin{figure}[t]
\centering
\includegraphics[width=0.95\columnwidth]{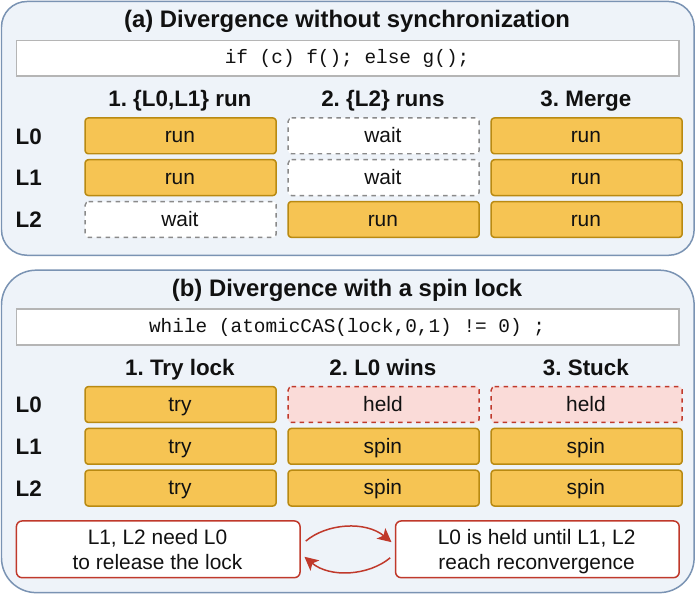}
\vspace{-2mm}
\caption{Ordinary divergence and spin-lock deadlock under strict IPDOM scheduling.}
\label{fig:simt-deadlock}
\vspace{-3mm}
\end{figure}

This fixed order becomes a problem when threads in the same warp
synchronize. Consider the spin lock in \cref{fig:simt-deadlock}. One
thread acquires the lock and leaves the acquisition loop, while the others
keep spinning. The stack keeps the spinning subgroup active and holds the
lock holder at the reconvergence point. The lock holder never releases the
lock, and the spinning threads never leave the loop. The result is a
SIMT-induced deadlock, even though the program would complete under a fair
MIMD schedule~\cite{habermaier2012simt,eltantawy2016mimd}. The challenge
is to let such subgroups make progress while keeping the compact control
state of stack-based SIMT.

Prior work addresses this problem at different costs. Static SIMT Deadlock
Elimination (SSDE) restructures synchronization code in the compiler, and
AWARE replaces strict stack traversal with split and reconvergence tables
and fairer scheduling~\cite{eltantawy2016mimd}. NVIDIA's Independent
Thread Scheduling (ITS) goes further and maintains execution state for
every thread~\cite{nvidia_volta_whitepaper,nvidia_cuda_its}. ITS removes
the ordering problem, but its control state grows with warp width.

Synchronization does not require this much independence. Threads waiting
on the same path can keep sharing one instruction stream. They only need a
different subgroup to run and end their wait. Stack-based reconvergence
already represents each subgroup as a thread mask and a program counter.
What it lacks is a way to schedule subgroups out of stack order.
Moreover, the number of subgroups that are live at the same time is
typically much smaller than the warp width, so this ability does not need
per-thread state.

Building on this observation, we present \textbf{\name{}}, a
compiler--hardware mechanism that makes divergent subgroups independently
schedulable while keeping stack-based reconvergence intact. A subgroup is
a thread mask and a resume program counter, so switching subgroups never
copies register state. The compiler inserts a yield instruction into loops
that may block. On a yield, the hardware rotates execution among one
pending subgroup per warp and a bounded pool of parked subgroups. The pool
capacity is sized to the workload, not to the number of threads. \name{}
targets SIMT-induced scheduling deadlocks, not deadlocks from an
application's own lock ordering.

We implement \name{}\footnote{\url{https://github.com/vortexgpgpu/vortex/tree/feature/simturn}\\
\url{https://github.com/vortexgpgpu/llvm/tree/simturn}} in the LLVM-based
compiler~\cite{jeong2025volt}, the SimX simulator, and the RTL of the
open-source Vortex GPGPU~\cite{vortex,vortex_first_2020}
(\cref{fig:SCS_TOP}, additions in red). We compare it against the IPDOM
baseline and a per-thread ITS implementation on the same platform. Across
warp widths from 4 to 32 threads, \name{} adds 17--24\% scheduler area
relative to IPDOM, compared with 39--91\% for ITS. On synchronization
kernels with independent work across subgroups, it tracks ITS and runs up
to 2.7$\times$ faster than software-serialized IPDOM. On control workloads
that never need cross-subgroup progress, it stays within 8\% of IPDOM's
runtime.

This paper makes the following contributions.
\begin{itemize}
    \item \textbf{Subgroup scheduling on stack-based reconvergence.}
    \name{} extends IPDOM with a yield instruction, one pending subgroup
    per warp, and a bounded pool of parked subgroups, without
    per-thread program counters.

    \item \textbf{An open-source implementation.} We implement \name{} in
    LLVM, SimX, and the Vortex RTL. The compiler inserts yields
    automatically, so kernels need no source changes.

    \item \textbf{A cost and progress evaluation.} We compare \name{}
    with IPDOM and ITS on synchronization and control workloads and on
    ASIC synthesis, and study its sensitivity to pool capacity and lock
    contention.
\end{itemize}

\section{Background and Related Work}
\label{sec:background}

\subsection{SIMT Divergence and Stack-Based Reconvergence}
\label{sec:background:simt}

SIMT processors execute scalar threads together on SIMD lanes, sharing one
instruction stream per warp and using an active-thread mask to select the
participating lanes. A branch whose outcome differs across lanes splits the
warp into divergent \emph{subgroups}. A subgroup is a set of threads in one
warp that follow the same control-flow path. The processor must preserve
each subgroup's resume PC and decide when the subgroups can resume
together~\cite{habermaier2012simt,eltantawy2016mimd}.
Stack-based control flow handles this with a reconvergence stack. Each
divergent entry records a program counter, a thread mask, and a
reconvergence point, which under IPDOM is the divergent branch's immediate
post-dominator. The scheduler runs one path while keeping the others as
suspended entries, and once the active path reaches its reconvergence
boundary another path can run, with masks combined once all participants
arrive. Strict stack traversal therefore imposes a depth-first order on
divergent paths~\cite{habermaier2012simt,eltantawy2016mimd}.

This arrangement shares fetch and control across active lanes and keeps
inactive paths in compact state, but it couples two distinct decisions,
where threads should reconverge and which subgroup may execute next, so
divergence management determines both SIMD utilization and the execution
order seen by communicating threads~\cite{arxiv:24:cf_management}.

Our chosen platform, Vortex, expresses this model with compiler-generated
control-flow instructions~\cite{vortex}. Divergent regions and
reconvergence are managed by \texttt{vx\_split} and \texttt{vx\_join},
while \texttt{vx\_pred} and \texttt{vx\_pred\_n} update thread participation
during divergent loops, and the core keeps one live program counter and
active mask per warp. This stack-based configuration is the IPDOM baseline
used in our evaluation.

\subsection{SIMT-Induced Deadlock}
\label{sec:background:deadlock}

A synchronization algorithm may require a waiting thread to leave other
threads opportunities to execute. Strict divergent-path serialization can
violate this requirement within a warp, because a suspended path cannot
run until the active path reaches its reconvergence point. Habermaier and Knapp analyze
this unfairness in stack-based SIMT execution~\cite{habermaier2012simt},
and ElTantawy and Aamodt characterize the resulting synchronization failures
as SIMT-induced deadlock~\cite{eltantawy2016mimd}.

\Cref{fig:simt-deadlock} contrasts ordinary divergence with a spin lock.
In \cref{fig:simt-deadlock}(a), serialization is harmless: each divergent
subgroup finishes on its own and the lanes merge. In
\cref{fig:simt-deadlock}(b), lane~0 acquires the lock and exits the
acquisition loop, while lanes~1 and~2 keep spinning. If the spinning path
stays active, IPDOM execution holds lane~0 at the loop's reconvergence
boundary. Lane~0 cannot release the lock until lanes~1 and~2 exit the
loop, and they cannot exit until it does, so a synchronization dependency
and a scheduling dependency form a cycle. The program could complete if
the lock holder ran, yet the SIMT execution order prevents it. NVIDIA
documents this class of intra-warp lock and mutex limitation for pre-Volta
execution~\cite{nvidia_volta_whitepaper,nvidia_cuda_its}.

The crucial distinction is between retaining a resume PC and making it
schedulable. The processor retains enough information to resume the lock
holder, but strict reconvergence ordering prevents it from doing so while
its sibling spins. Scheduling the lock holder before the spinning subgroup
reconverges can break this cycle without assigning a separate program
counter to every waiting lane.

SIMT-induced deadlock is distinct from a cycle in the application's lock
acquisition order. It is also distinct from a dependency on a workgroup
that has not been scheduled onto the device, which GPU workgroup progress models address~\cite{sorensen2021progress}. The mechanism studied here
addresses scheduling dependencies among divergent subgroups. Its progress
argument depends on reaching suitable yield points, retaining the required
parked subgroups, and allowing the participating resident work and memory
operations to advance. These conditions separate the execution mechanism
from the correctness of the synchronization protocol itself.

\subsection{Approaches to Forward Progress}
\label{sec:background:progress}

\begin{figure}[t]
\centering
\includegraphics[width=0.95\columnwidth]{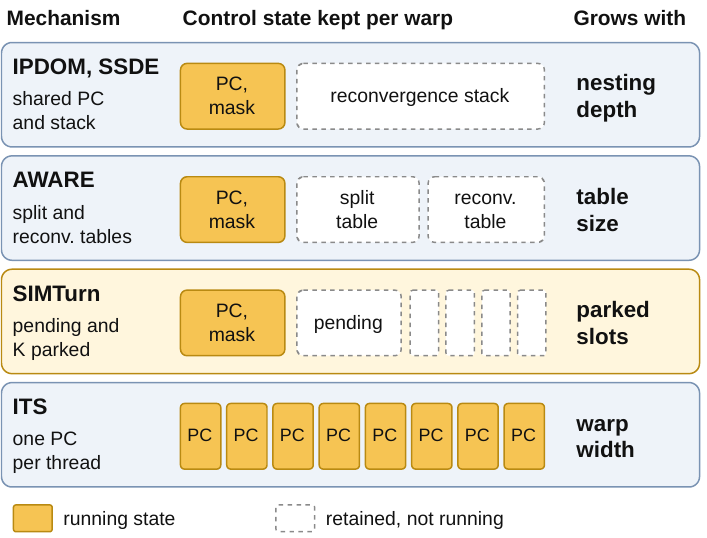}
\vspace{-2mm}
\caption{Per-warp control state across various SIMT control-flow divergence mechanisms.}
\label{fig:control-state}
\vspace{-3mm}
\end{figure}

\Cref{tab:simt-progress} compares the principal mechanisms by their
execution state and how they expose opportunities for progress. The table
summarizes mechanisms rather than assigning unconditional deadlock-freedom
guarantees, which depend on the supported programs and scheduling model.
\Cref{fig:control-state} schematically compares per-warp control state
with warp width $W$ and $K$ parked slots. SSDE retains
IPDOM's hardware state, and AWARE replaces the stack with split and
reconvergence tables. \name{} adds one pending slot and $K$ parked slots
to the stack, while ITS maintains a PC per thread.

\begin{table*}[t]
\centering
\footnotesize
\caption{Control-flow mechanisms relevant to intra-warp progress.}
\begin{tabular}{@{}p{0.12\textwidth} p{0.22\textwidth} p{0.28\textwidth} p{0.29\textwidth}@{}}
\toprule
\textbf{Mechanism} & \textbf{Control-flow state} &
\textbf{Execution policy} & \textbf{Scope and requirements} \\
\midrule
IPDOM~\cite{habermaier2012simt} &
Shared warp PC and reconvergence stack &
Strict stack traversal serializes divergent paths &
Compact control; inactive siblings can be prevented from satisfying a wait \\
\midrule
SSDE~\cite{eltantawy2016mimd} &
Existing stack-based state &
Compiler restructures synchronization to expose path progress &
Depends on static detection and CFG transformation \\
\midrule
AWARE~\cite{eltantawy2016mimd} &
Split and reconvergence tables; scheduling FIFO &
Fairer split scheduling with delayed reconvergence &
Decouples scheduling from stack order; table capacity and virtualization matter \\
\midrule
DWS~\cite{meng2010dws} &
Sub-warp execution contexts &
Subdivides warps to tolerate branch and memory divergence &
Targets execution efficiency rather than synchronization progress directly \\
\midrule
Hanoi~\cite{arxiv:24:cf_management} &
Split/reconvergence stacks and convergence state &
Models explicit convergence operations and sibling-path yielding &
Examines control-flow behavior inferred from modern NVIDIA execution \\
\midrule
ITS~\cite{nvidia_volta_whitepaper,diamos2016convergence} &
Per-thread execution state and convergence state &
Threads can advance independently and regroup for shared execution &
Supports finer-grained scheduling; requires thread-level control state \\
\midrule
\name{} &
Existing stack plus pending and parked subgroups &
Compiler-directed yielding rotates execution among subgroups &
Requires yield coverage and sufficient parked capacity \\
\bottomrule
\end{tabular}
\label{tab:simt-progress}
\end{table*}

\paragraph{Compiler restructuring and split scheduling.}
SSDE identifies potentially unsafe synchronization and transforms the
control-flow graph to let otherwise suspended paths make
progress~\cite{eltantawy2016mimd}. It preserves the stack hardware,
placing the responsibility for exposing progress in the compiler.
AWARE instead separates split tracking from reconvergence through a
Split Table, a Reconvergence Table, and a scheduling FIFO. Its scheduling
and delayed-reconvergence mechanisms relax depth-first execution. The
work also explores memory virtualization of these tables. AWARE is the
closest architectural precedent for \name{}. Both schedule divergent
splits, but \name{} integrates explicit yielding and bounded parked
subgroups with the existing stack-based mechanism. The comparison is thus
about the machinery and software interface needed to expose progress,
not the discovery that subgroup scheduling helps synchronization.

\paragraph{Independent thread scheduling.}
Beginning with Volta, NVIDIA maintains execution state at per-thread
granularity, including program counters and call stacks, and dynamically
groups compatible threads for shared SIMT
execution~\cite{nvidia_volta_whitepaper,nvidia_cuda_its}. This permits
threads to yield independently and supports synchronization patterns
that can deadlock under strict shared-PC execution. The convergence-barrier
model described by Diamos et al. includes thread states, resume PCs,
participation masks, and explicit convergence
operations~\cite{diamos2016convergence}. These descriptions motivate the
per-thread design point evaluated in this paper. Our ITS baseline is an
implementation of this design point on the same GPU platform.

\paragraph{Busy-wait overheads.}
Busy-wait synchronization that does complete can still waste resources.
ElTantawy and Aamodt detect spin loops at run time (DDOS) and back off
spinning warps (BOWS) to reduce the issue slots and memory traffic they
consume~\cite{eltantawy2018bows}. Other work reduces synchronization cost
through synchronization-aware warp scheduling~\cite{liu2015saws} and by
moving fine-grained locks into scratchpad memory~\cite{wang2019fast}.
These approaches target contention between warps, whereas \name{} addresses scheduling dependencies among subgroups
inside one warp. Run-time spin detection could also help \name{} find
yield points.

\paragraph{Divergence efficiency and control-flow flexibility.}
Dynamic Warp Formation regroups threads from different warps at common
PCs to improve SIMD utilization~\cite{fung2007dwf}, and Thread Block
Compaction applies a similar regrouping at reconvergence points within a
thread block~\cite{hpca:11:tbc}, while Dynamic Warp Subdivision addresses
branch and memory divergence through independently executing
sub-warps~\cite{meng2010dws}. SIMD Re-convergence at Thread Frontiers
generalizes reconvergence-point detection beyond the immediate
post-dominator to handle unstructured control flow~\cite{micro:11:thread_frontiers},
and Dual-Path execution extends each stack entry to hold two live paths
that can be scheduled interchangeably~\cite{hpca:13:dual-path}. GPU
Subwarp Interleaving explores fine-grained interleaving at a subwarp
granularity similar to \name{}, though its authors report that the added
cost did not justify the benefit in a commercial
design~\cite{hpca:22:subwarp}. Hanoi studies control-flow management
using inferred modern NVIDIA ISA behavior, including explicit convergence
and yielding operations~\cite{arxiv:24:cf_management}.

Related work improves divergence behavior through compaction
prediction~\cite{isca:12:capri}, divergence analysis and CFG
restructuring~\cite{pact:11:divergence,hpcs:16:cf_restructuring,cc:14:taming},
control-flow melding~\cite{cgo:22:dram}, and multi-path
execution~\cite{hpca:14:multipath}. Other studies explore the divergence
management design space~\cite{micro:14:exploringDSE}, characterize
unstructured control flow~\cite{caches:11:characterization}, or address
memory-system and application-level effects through divergence-aware
caching and herding~\cite{ics:15:dacache,pact:12:data_herding}.
These techniques target the efficiency of divergent execution, whereas
\name{} targets forward progress between divergent subgroups.


\paragraph{Positioning \name{}.}
\name{} sits between compiler-only restructuring and per-thread
scheduling. Unlike SSDE, it leaves synchronization code unchanged and only
marks where a waiting path may yield. Unlike AWARE, it keeps the
reconvergence stack for ordinary divergence and uses its added state only
when a subgroup yields. Unlike ITS, this state
scales with the number of parked subgroups, not with the number of
threads. In return, it requires yield coverage and sufficient parked
capacity.

\section{Motivation and Design Principles}
\label{sec:motivation}

\subsection{Challenges of Deadlock-Free SIMT Execution}
\label{sec:motivation:challenge}

Stack-based reconvergence keeps control state compact, but its fixed path
order can leave a lock holder unschedulable while its siblings spin. Our HT
and LCList kernels (\cref{tab:kernels}) run into exactly this problem. Our
Vortex ITS baseline shows what it costs to lift that order with per-thread
scheduling. It adds per-thread PCs and intra-warp scheduling state on top
of the existing mechanism, which is far more than a change in scheduling
policy alone.

\begin{figure*}
    \centering
    \includegraphics[width=\textwidth]{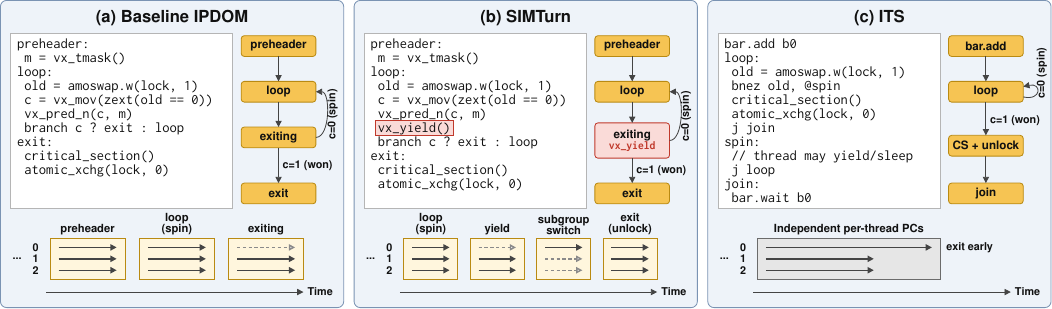}
    \caption{Baseline IPDOM, \name{}, and ITS execution of the
    HT bucket-lock spin loop.}
    \label{fig:lockht-comparison}
\end{figure*}

\Cref{fig:lockht-comparison} compares the three designs on the HT
bucket-lock spin loop. The code and control-flow panels show the compiler
instrumentation, where the highlighted \texttt{vx\_yield} in panel~(b) is
the only instruction that \name{} adds. The timelines show that IPDOM
stalls once the winner is masked off, while a yield lets the lock holder
run.

A fully independent program counter for every thread is sufficient for
forward progress, but it is not necessary. IPDOM already separates a warp
into divergent subgroups, each with an active mask and a suspended
control-flow location. Its only limitation is that these subgroups are
consumed in a fixed depth-first order. A subgroup that could release a
lock or another shared resource is therefore already represented by the
divergence mechanism, yet it stays unschedulable while another subgroup
waits on it.

A second challenge follows. Subgroup scheduling must not recreate the cost
that \name{} seeks to avoid. Arbitrary divergence can in theory create as
many distinct subgroups as there are threads in a warp, and provisioning
for that worst case would again approach ITS-like state.

\subsection{Design Principles of \name{}}
\label{sec:motivation:principles}

Four principles guide how \name{} divides forward progress between the
compiler, ISA, and microarchitecture.

\textbf{Schedule divergence, not individual threads.}
The unit that needs independent progress is usually a divergent subgroup,
not a single lane. Threads on the same path already share a PC and run
efficiently in SIMT fashion. \name{} therefore attaches schedulability to
subgroups. Per-lane algorithmic state stays in the existing architectural
registers. The only control-flow state needed to resume a subgroup is its
thread mask and resume PC, so switching between subgroups never copies
register state.

\textbf{Preserve the IPDOM fast path.}
Forward-progress support should not penalize the common case where
ordinary divergence terminates normally. \name{} builds on Vortex's
existing split/join model and does not replace it. \texttt{vx\_split} and
\texttt{vx\_join} continue to identify divergence and reconvergence
regions. Subgroup scheduling activates only when execution needs to
transfer between paths, so converged execution and non-blocking divergence
keep the baseline's simple warp-level behavior.

\textbf{Bound expensive resident state.}
\name{} keeps divergence state in dense, inexpensive storage instead of
large arrays of timing-critical per-thread state. Parked subgroups live in
a bounded pool whose capacity is a configurable architectural parameter
and whose bulk storage is BRAM-backed on FPGA implementations. The
compiler helps by structuring blocking regions so that a blocking
operation needs only a few simultaneously live subgroups. Resident
hardware cost therefore scales with the configured pool capacity and not
with thread-granular state for every lane.

\textbf{Use software to expose progress, not scheduling resources.}
The compiler is better positioned to identify where execution may wait,
while hardware is better positioned to choose which ready subgroup runs
next. LLVM identifies potentially blocking loops and inserts a lightweight
\texttt{vx\_yield} at appropriate loop backedges. The hardware interprets
this instruction by transferring execution to another runnable subgroup.
Unlike an ITS-style convergence optimizer, the compiler does not allocate
named hardware barriers or manage their lifetimes across nested control
flow. This keeps the compiler transformation simple and leaves scheduling
policy in the microarchitecture.

Together, these principles retain SIMT execution wherever threads
naturally move together and introduce independence only between the
divergent subgroups that require it. The added control-flow state stays
proportional to the number of resident subgroups, not the number of
threads. The next section describes how these principles are realized in
the Vortex microarchitecture and the LLVM divergence pipeline.

\section{\name{} Design and Implementation}
\label{sec:framework}

\name{} combines a compiler-generated scheduling point with hardware
that retains and selects divergent subgroups. We first explain this
execution model, then describe the scheduler and the compiler
transformation. We also separate the conditions for correct subgroup
progress from the pipeline mechanisms used to implement the design.

\subsection{Execution Model and ISA Interface}
\label{sec:framework:execution}

A subgroup consists of a thread mask and a resume PC. The mask determines
which lanes participate when the subgroup resumes, while per-thread
operands and algorithmic state remain in the existing registers. \name{}
does not replicate the IPDOM stack for each subgroup. The existing stack
continues to maintain the warp's ordinary reconvergence state, including
its return/reconvergence PC, next PC, and active mask. \name{} adds two
forms of resumable state per warp, a \emph{pending subgroup} and a pool of
\emph{parked subgroups}. Each parked subgroup is stored as only a mask and
a resume PC. Pending work is captured by a divergent loop predicate and
can either be restored through ordinary loop completion or selected for
independent execution. Parked work has been descheduled and retained for
later selection.

\begin{figure}[!ht]
\centering
\includegraphics[width=0.95\columnwidth]{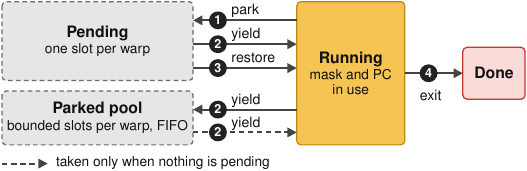}
\vspace{-2mm}
\caption{Lifecycle of a subgroup within a warp.}
\label{fig:lifecycle}
\vspace{-3mm}
\end{figure}

\begin{figure*}
    \centering
    \includegraphics[width=0.95\linewidth]{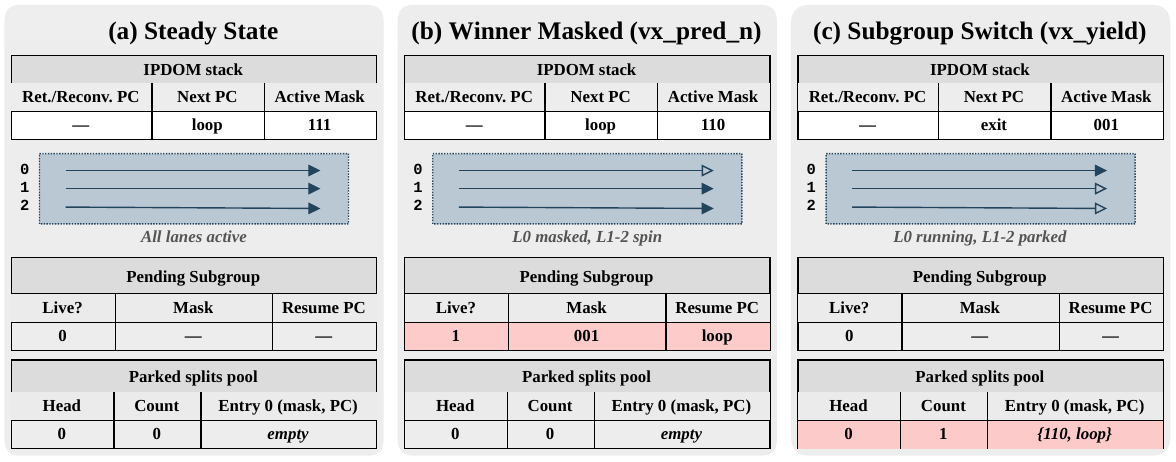}
    \caption{Subgroup-state transitions during HT lock acquisition.}
    \label{fig:threadsplit-pool}
\end{figure*}

\Cref{fig:lifecycle} shows how a subgroup moves between these states. It
is running while its mask and PC are in use, pending or parked while it
is retained, and done once its lanes exit.
\Cref{fig:threadsplit-pool} follows these states through the HT lock
acquisition. In panel~(a), all three lanes are active and no additional
subgroup is needed. In panel~(b), lane~0 acquires the lock, and
\texttt{vx\_pred\_n} leaves lanes~1--2 active while it captures lane~0 in
the pending register. In panel~(c), the yield selects the pending winner,
saves the spinning subgroup in the pool, and changes the executing mask to
lane~0. The winner can now enter the critical section and release the
lock. Later subgroup selections let the parked lanes resume at their
waiting PC and retry the acquisition. The \texttt{loop} and \texttt{exit}
labels in the figure mark control-flow phases, not individual
instructions.

The ISA addition is \texttt{vx\_yield}, a warp-control instruction with
no register operands. It exposes a point at which the current subgroup can
be descheduled in favor of another retained subgroup. The instruction
is handled as a warp-stalling control operation, so instruction execution
waits until the scheduler has resolved the transition. Its purpose is
scheduling, and any memory ordering that a synchronization protocol
requires stays part of that protocol.

\subsection{Subgroup Scheduler}
\label{sec:framework:scheduler}

Each numbered event in \cref{fig:lifecycle} corresponds to one handler of
the scheduler. We describe them in the same order.

\paragraph{State organization.}
The pending register holds a valid bit, a thread mask, and a resume
PC. The parked pool holds mask/PC pairs with a head pointer and an
occupancy counter. Two bookkeeping masks track completed lanes and
lanes represented in the pool. They prevent a later restoration from
reactivating completed work or duplicating lanes owned by another
schedulable subgroup. Let $K$ denote the number of parked slots per warp.
The active and pending subgroups are accounted for separately.

\paragraph{Capturing pending work (event 1, park).}
A divergent \texttt{vx\_pred} or \texttt{vx\_pred\_n} already computes
masks for lanes remaining in the loop and lanes leaving it. When both
sets are nonempty, the warp-control unit raises
\emph{pred\_park\_valid} with the exiting mask. The scheduler records
these lanes as pending while the remaining lanes continue. Lanes that
leave through the same resume PC can accumulate in the pending mask across
iterations. This compatibility requirement is essential, because mask
accumulation must not combine lanes that need different resume PCs.
The recorded PC is taken from the warp's post-increment PC state.

\paragraph{Selecting another subgroup (event 2, yield).}
The scheduler responds to \emph{yield\_valid} by saving the current
subgroup in the pool and then selecting another subgroup to run. Pending
work has priority and can be installed directly, as in
\cref{fig:threadsplit-pool}(c). Otherwise, execution rotates to the oldest
parked subgroup. The FIFO orders parked subgroups, and pending priority is
a separate part of the policy. After a switch, lanes that were previously
prevented from executing can run and satisfy a sibling's wait.

\paragraph{Restoration (event 3, restore).}
When no lanes remain on the loop's continuing path, the warp-control unit
raises \emph{pred\_restore\_valid}. The scheduler merges a pending mask
back into the running subgroup only when the pending resume PC matches
the restoration point. This check prevents restoration from applying an
older mask snapshot to unrelated subgroup work. When no compatible pending
subgroup exists, the running subgroup remains the restoration basis, and
the preheader mask in \texttt{rs2} is not used.

\paragraph{Completion (event 4, exit).}
A kernel-exiting mask-clear operation records the completed lanes while
subgroup work remains and selects pending work or the pool head. The warp
retires only when no retained work survives. Temporary mask narrowing in
the runtime must keep its existing behavior, because temporarily inactive
lanes are not equivalent to permanently completed lanes.

\Cref{alg:scs-scheduler} summarizes the four event handlers and their
selection policy. It abstracts the physical queue operations and the mask
bookkeeping, and it is not a complete specification of arbitrary nested
control flow. The RTL additionally updates the \texttt{done} and
\texttt{in\_pool} masks when subgroups are inserted, removed, completed,
and restored.

\begin{algorithm}[t]
\small
\DontPrintSemicolon
\SetAlgoNoLine
\SetAlgoVlined
\LinesNotNumbered
\SetKwInput{KwIn}{Input}
\SetKwProg{ScsProc}{Procedure}{:}{}
\SetKw{ScsReturn}{return}
\caption{\small Warp Scheduler Subgroup Management}
\label{alg:scs-scheduler}
\KwIn{Per-warp: pending $\langle$live, mask, PC$\rangle$; pool
(head, count); done, in-pool masks; live mask; warp PC}

\ScsProc{\textnormal{\textsc{InstallNext}}$()$}{
  \eIf{pending.live}{
    live $\gets$ pending.mask $\land \lnot$done;\ \
    PC $\gets$ pending.PC;\ pending.live $\gets$ \textbf{false}\;
  }{
    \eIf{pool.count $>$ 0}{
      (mask, PC$'$) $\gets$ pop pool head;\ \
      live $\gets$ mask $\land \lnot$done;\ PC $\gets$ PC$'$\;
    }{ \ScsReturn \textbf{false} \tcp*[l]{nothing runnable}}
  }
  \ScsReturn \textbf{true}\;
}

\ScsProc{\textnormal{\textsc{OnPredPark}}$(\mathit{else\_mask})$ \textnormal{(event 1)}}{
  pending.mask $\gets$ pending.mask $\lor \mathit{else\_mask}$;\ \
  pending.PC $\gets$ warp.PC;\ pending.live $\gets$ \textbf{true}\;
}

\ScsProc{\textnormal{\textsc{OnYield}}$()$ \textnormal{(event 2)}}{
  push (live, warp.PC) to pool tail\;
  \textsc{InstallNext}$()$ \tcp*[l]{no-op if nothing runnable}
}

\ScsProc{\textnormal{\textsc{OnPredRestore}}$()$ \textnormal{(event 3)}}{
  \If{pending.live \textbf{and} pending.PC $=$ warp.PC}{
    live $\gets$ live $\lor$ pending.mask;\ pending.live $\gets$ \textbf{false}\;
  }
}

\ScsProc{\textnormal{\textsc{OnTmcExit}}$(\mathit{exiting\_mask})$ \textnormal{(event 4)}}{
  \If{pending.live \textbf{or} pool.count $>$ 0}{
    done $\gets$ done $\lor \mathit{exiting\_mask}$\;
  }
  \If{\textbf{not} \textsc{InstallNext}$()$}{retire warp\;}
}
\end{algorithm}

\subsection{Compiler Transformation}
\label{sec:framework:llvm}

The compiler exposes waiting points through the existing branch-divergence
pass. That pass already inserts \texttt{vx\_pred} or \texttt{vx\_pred\_n} at
each conditional loop exit to identify the lanes that leave a loop, and
the hardware uses the same event to capture a pending subgroup. \name{} adds a
yield after the predicate in loops that may block, as highlighted in
\cref{fig:lockht-comparison}. For each divergent loop, innermost first,
the pass does the following.
\begin{enumerate}
    \item Decide whether the loop may block, using the \textsc{MayBlock}
    classifier described below.
    \item Insert \texttt{vx\_tmask} in the loop preheader to snapshot the
    entry mask.
    \item For each conditional exiting branch, compute its predicate with
    \texttt{vx\_mov} and insert it immediately before the branch.
    \item If the branch leaves the loop through its first successor, insert
    \texttt{vx\_pred\_n} before the branch and, when the loop may block, a
    \texttt{vx\_yield} right after it. Otherwise, insert \texttt{vx\_pred}
    and no yield.
    \item Replace the branch condition with the new predicate.
\end{enumerate}

\paragraph{Identifying candidate loops.}
The current \textsc{MayBlock} classifier recognizes native
\texttt{atomicrmw} and \texttt{cmpxchg} instructions, inline assembly
containing \texttt{amo} or \texttt{lr.}, and direct calls with
\texttt{atomic} in the callee name. These cases cover the atomic forms
targeted by the implementation. The classifier identifies candidate
waiting loops and does not prove that an operation will block. It does not
comprehensively detect load-only polling, indirect calls, or
synchronization hidden behind unrecognized interfaces.

\paragraph{Placement and ordering.}
The transformation runs after control-flow structurization, which
provides the divergent-loop representation with its preheaders and
conditional exits. Yield coverage additionally depends on the placement of
the relevant exiting branch. The steps above insert \texttt{vx\_yield} for
the exit-on-\texttt{succ\_0} form lowered with \texttt{vx\_pred\_n}. The
alternative branch form retains predicate insertion without a yield. The
transformation therefore covers this specific insertion case and does not
give a coverage guarantee that is independent of branch orientation.

The yield follows the predicate so that the winner has been captured before
the remaining lanes request a switch. It is emitted as side-effecting,
memory-clobbering inline assembly to constrain compiler motion. This
compiler constraint is distinct from hardware memory-ordering semantics.
Loops not selected for yield insertion retain their existing control-flow
instrumentation, and a compiler flag disables yield insertion for ablation.

This keeps the interface narrow. Software identifies where a waiting
path should offer a scheduling opportunity, while hardware retains and
selects the subgroups. The compiler does not name, allocate, or reclaim
scheduling state.

\subsection{Correctness Requirements and Progress Scope}
\label{sec:framework:progress}

Subgroup switching must preserve execution before it can improve progress.
Three requirements govern the state transitions. First, every unfinished
lane must retain a valid resume PC, in the executing subgroup, the pending
register, the parked pool, or the ordinary divergence state. Second,
scheduling ownership must remain consistent. Restoring a mask must not
duplicate a lane already assigned to another schedulable subgroup or
reactivate a completed lane. Third, retaining the IPDOM stack requires
respecting its reconvergence obligations across subgroup switches. The
simple predicate-based loop in \cref{fig:threadsplit-pool} illustrates
capture and selection. It does not by itself establish correctness for
all possible nested \texttt{vx\_split}/\texttt{vx\_join} interleavings.

The progress rationale is that a waiting subgroup offers an execution
opportunity to a subgroup that can satisfy its dependency. This guarantee
holds only when all five of the following conditions are met.
\begin{enumerate}
    \item \textbf{Yield coverage.} Every potentially blocking loop on the
    path carries a \texttt{vx\_yield}.
    \item \textbf{Sufficient capacity.} The subgroups waiting at any
    time never exceed the pool capacity $K$.
    \item \textbf{Fair selection.} The scheduler eventually selects every
    pending or parked subgroup.
    \item \textbf{Resident dependent work.} The subgroup that can satisfy
    a wait is already resident on the device.
    \item \textbf{Acyclic application dependencies.} The program's own
    synchronization has no circular lock order.
\end{enumerate}
An execution opportunity does not guarantee that a particular lane wins
every contended acquisition.

The resident pool makes capacity part of these conditions. When demand
exceeds $K$, the bounded implementation cannot claim the same progress
behavior without an additional capacity-handling mechanism, so we measure
capacity sensitivity explicitly. Likewise, the atomic-loop
classifier bounds the compiler coverage. Arbitrary nested divergence with
independently interleaved stack state is outside the scope of the
evaluated implementation.

\subsection{Microarchitecture Implementation}
\label{sec:framework:microarch}

\paragraph{Pipeline integration.}
\Cref{fig:SCS_TOP} locates the additions in the GPU pipeline. The
implementation extends warp control and subgroup scheduling while
retaining the existing IPDOM-stack structure. The new yield instruction
uses the Vortex custom warp-control encoding (\texttt{funct7}=5,
\texttt{funct3}=0). The predicate masks are reused to generate the park
and restore events, and the decoded instruction generates the yield
event. These signals pass through the existing registered warp-control
interface, which avoids a separate subgroup writeback or commit path.

\paragraph{Pool storage.}
The parked pool is the bulk of the additional state and maps to dual-port
BRAM on an FPGA. With $N_w$ warps, $W$ lanes per warp, PC width $P$, and
$K$ parked slots per warp, its payload storage is
\[
    S_{\mathrm{pool}} = N_w K(W+P)\ \text{bits},
\]
excluding pending registers, pointers, counters, and bookkeeping masks.
This accounting separates the configurable number of parked slots from
the width-dependent size of each mask. The pool is addressed by
$\{\mathit{wid},\mathit{slot}\}$. At most one subgroup-control event
retires per cycle, which allows a shared one-write/one-read organization.

\begin{figure}[!ht]
\centering
\includegraphics[width=0.95\columnwidth]{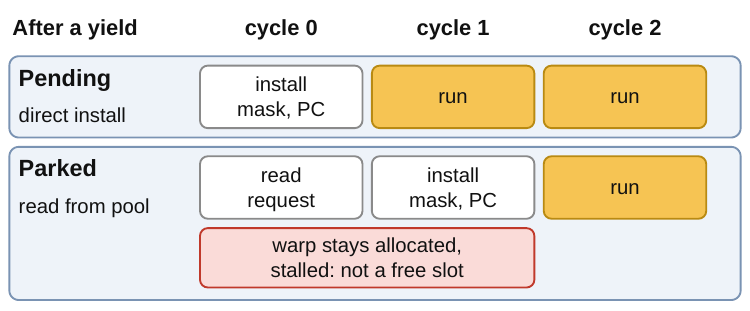}
\vspace{-2mm}
\caption{Pending and parked activation after \texttt{vx\_yield}.}
\label{fig:activation}
\vspace{-3mm}
\end{figure}

\paragraph{Pipelined activation.}
A combinational push/pop/select sequence can place pool-memory access
on the next-PC path. The implementation therefore pipelines pool
activation. A read request selects a parked entry, and its mask/PC pair is
installed one cycle later. A pending subgroup can instead be installed
directly, which avoids the pool-read latency. This difference between
direct pending selection and pipelined pool selection follows from the two
subgroup classes shown in \cref{fig:threadsplit-pool}.
\Cref{fig:activation} compares the two paths.

During a pipelined activation, the warp remains active but stalled.
The CTA dispatcher must not interpret this interval as a free warp slot.
Only true completion, with no pending or parked work and no activation
in flight, permits retirement and reuse. Preserving ownership across this
interval prevents a newly dispatched task from overwriting a subgroup
that has been selected but not yet installed.

\section{Evaluation}
\label{sec:evaluation}

\paragraph{Experimental setup.}
We evaluate our implemented \name{} design using Vortex RTL simulation~\cite{vortex}, based on Verilator~\cite{verilator}.
The RV32IMAFC GPU has one core, eight resident warps, and $W\in\{4,8,16,32\}$ threads per warp.
SIMD width scales with $W$, while issue width remains one.
Each configuration has 16\,KiB, four-way instruction and data caches, with no L2 cache enabled.
OpenCL kernels are compiled at \texttt{-O3} using LLVM-Vortex 20.1.8 and launched through
PoCL~\cite{paper:llvm,paper:pocl:2015}.

We compare three configurations. For the IPDOM baseline, we convert the
naturally blocking code to a software-serialized version
(\texttt{IPDOM\_SW}). On SIMTurn, we run the same natural code with
compiler-inserted yields. On ITS, we run it unmodified, because per-thread
scheduling needs no yield to make progress. \texttt{IPDOM\_SW} executes
complete synchronization operations in local-work-item order, so only one
lane reaches a blocking acquisition at a time. The lock protocol and
inputs are otherwise identical.

\paragraph{Measurements and correctness.}
We report simulated device cycles, dynamic warp instructions, and average SIMD utilization. Device measurements include kernel dispatch and device-side runtime instructions, but exclude
host compilation and transfers. Time and instruction counts are normalized to \texttt{IPDOM\_SW} for synchronization and control workloads. Host checks validate final data and released locks, while order-dependent distance updates are checked by CPU replay of the recorded serialization.

\paragraph{Evaluated benchmarks.}
We evaluate 9 synchronization and 4 control workloads, summarized in
\cref{tab:kernels}. Our OpenCL synchronization kernels follow the patterns
evaluated by AWARE~\cite{eltantawy2016mimd}, with explicitly parameterized
contention~\cite{maucher2025atomics}. The controls are Vortex's
existing atomic-reduction and histogram microbenchmarks, a rank/counting
sort, and the Rodinia pathfinder kernel~\cite{vortex_github,benchmark:rodinia}.
They exercise atomics, loops, and divergence without requiring
cross-subgroup progress, so they isolate any overhead that \name{} adds
when forward progress was never at stake.

\begin{table}
\footnotesize
\centering
\caption{Evaluated benchmarks.}
\label{tab:kernels}
\begin{tabular}{@{}p{0.15\linewidth} p{0.47\linewidth} p{0.27\linewidth}@{}}
\toprule
\textbf{Kernel} & \textbf{Description} & \textbf{Sweep parameters} \\
\midrule
\multicolumn{3}{@{}c}{\textit{\textbf{Synchronization}}} \\
TL-CG/FG & Protected-counter increment; CG: one lock, FG: independent locks; adjacent lanes share targets & 256 work-items; FG: 2/8/32 locks \\
AM-CG/FG & Protected-maximum update, same CG/FG split & 256 work-items; 25/50/100\% participation \\
HT & Preallocated chained hash-table insertion~\cite{herlihy2020multiprocessor} & 128/512 nodes, 1/8/64 buckets \\
LCList & Hand-over-hand sorted-list search \& per-key update~\cite{gramoli2015synchrobench,synchrobench_code} & 128 ops, 8/32/128 keys, 0/50/100\% updates \\
ATM & Two-account transfer, locks acquired in account order & 2/8/32 accounts \\
CP-DS & Endpoint-locked distance-constraint projection on a perturbed grid~\cite{brownsword2009cloth} & 512 constraints $\times$2, $4^2$/$8^2$/$16^2$ grid \\
BH-ST & Octree offset propagation; child polls parent's publication~\cite{barnes1986hierarchical,coplin2015source} & 128/256/512 nodes \\
\midrule
\multicolumn{3}{@{}c}{\textit{\textbf{Control}}} \\
AtomicRed & Max-contention reduction: every lane \texttt{atomic\_add}s into one global counter & $n{=}1024$ \\
Histogram & Per-bin \texttt{atomic\_add}; contention only within a bin & $n{=}1024$ \\
Rank Sort & Counting/rank sort; all-pairs comparison, no atomics or locks & $n{=}256$ \\
Pathfinder & Rodinia dynamic-programming wavefront, halo-cached in local memory~\cite{benchmark:rodinia} & $16{\times}64$ grid, $y{=}2$, block $16$ \\
\bottomrule
\end{tabular}
\end{table}

\subsection{Synchronization Microbenchmarks}
\label{sec:eval:microbenchmarks}

\begin{figure*}
    \centering
    \begin{subfigure}[b]{0.32\linewidth}
        \centering
        \includegraphics[width=\linewidth]{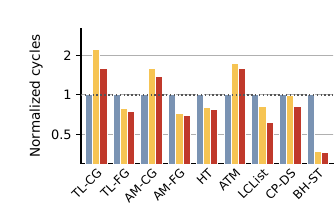}
        \caption{Normalized device cycles.}
        \label{fig:microbench-time}
    \end{subfigure}
    \hfill
    \begin{subfigure}[b]{0.32\linewidth}
        \centering
        \includegraphics[width=\linewidth]{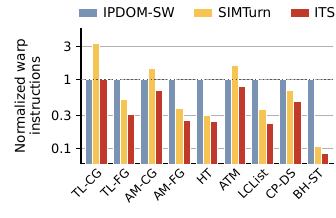}
        \caption{Normalized dynamic warp instructions.}
        \label{fig:microbench-instructions}
    \end{subfigure}
    \hfill
    \begin{subfigure}[b]{0.32\linewidth}
        \centering
        \includegraphics[width=\linewidth]{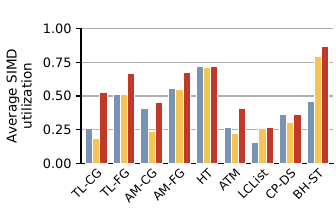}
        \caption{Average SIMD utilization.}
        \label{fig:microbench-utilization}
    \end{subfigure}
    \caption{Synchronization microbenchmark performance.}
    \label{fig:microbench}
\end{figure*}

\Cref{fig:microbench} reports geometric means over each kernel's parameter
sweep, averaged across all four warp widths. SIMTurn tracks \texttt{ITS}
closely on most synchronization benchmarks. The bounded pool recovers
essentially the same outcome as full per-thread scheduling, without paying
for it.

The kernels fall into two groups. Where genuine independent parallelism
exists across locks or lanes (TL-FG, AM-FG, HT, LCList, and BH-ST),
SIMTurn and ITS both beat \texttt{IPDOM\_SW}. The largest gain is on
BH-ST, where SIMTurn is $2.7\times$ faster and SIMD utilization rises from
$0.46$ to $0.79$. On CP-DS, SIMTurn executes fewer warp instructions but
only matches \texttt{IPDOM\_SW}'s cycles. The second group, TL-CG, AM-CG,
and ATM, has a single point of contention that every lane must pass
through. Here SIMTurn and ITS are both \emph{slower} than the plain retry
loop (SIMTurn $2.2\times$ on TL-CG). With no independent subgroup to
schedule to, yields and switches are pure overhead on top of a
serialization that \texttt{IPDOM\_SW} already achieves for free.

\Cref{fig:microbench}(\subref{fig:microbench-instructions}) and
\cref{fig:microbench}(\subref{fig:microbench-utilization}) show the same
split. In the first group, SIMTurn and ITS execute far fewer warp
instructions than \texttt{IPDOM\_SW}, down to $0.11\times$ and
$0.09\times$ on BH-ST. In the second group, SIMTurn executes more, up to
$3.3\times$ on TL-CG, while ITS executes no more than \texttt{IPDOM\_SW}
($0.69$--$0.99\times$). SIMTurn's SIMD utilization rises sharply on LCList
(from $0.16$ to $0.26$) and BH-ST, stays level on the other kernels of the first group, and falls
in the second group, from $0.26$ to $0.19$ on TL-CG. ITS reaches the
highest utilization on seven of the nine kernels, including $0.52$ on
TL-CG.

\subsection{Control Workloads}
\label{sec:eval:control}

\begin{figure}
    \centering
    \includegraphics[width=\linewidth]{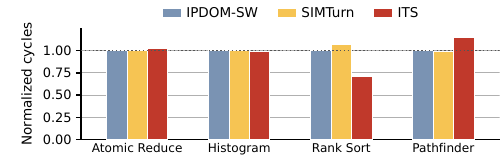}
    \caption{Control workloads performance normalized to
    \texttt{IPDOM\_SW} device cycles.}
    \label{fig:control-overhead}
\end{figure}

\Cref{fig:control-overhead} confirms that the IPDOM fast path is mostly
untouched on all four kernels. Averaged across all four warp widths,
SIMTurn stays within $8\%$ of \texttt{IPDOM\_SW}. These kernels never need
cross-subgroup progress, so the split/pred machinery never fires as
anything more than ordinary divergence. ITS deviates further in both
directions. It runs Rank Sort in $0.72\times$ the cycles of
\texttt{IPDOM\_SW}, with $0.77\times$ the warp instructions, and
Pathfinder in $1.15\times$.

\subsection{Case Studies}
\label{sec:eval:casestudies}

We now look inside the scheduler itself, at fixed $W{=}32$, to ask two
questions that are central to \name{}'s design. \emph{How many schedulable
subgroups does a divergent kernel actually need at once?} \emph{How does
lock contention change what the scheduler is doing?} We instrument the RTL
to additionally emit, per warp-cycle, the number of \emph{live} subgroups
(the running subgroup plus its warp's pending entry and pool occupancy) and
the number of \emph{runnable} subgroups not currently executing (pending $+$ pool). This instrumentation is not otherwise part of the
evaluated design.

\paragraph{Case A: How many subgroups does SIMTurn actually use?}
We measure the three synchronization kernels with the widest range of
lock-dependency structure. These are HT (8 buckets, 256 work-items), ATM
(8 accounts, 256 work-items, 4 rounds), and LCList (32 keys, 128
work-items, 50\% updates). We run them at SIMTurn's default capacity
($K{=}W{=}32$) and across $K\in\{1,2,4,8,16\}$.

\begin{figure}
    \centering
    \begin{subfigure}[b]{0.49\linewidth}
        \centering
        \includegraphics[width=\linewidth]{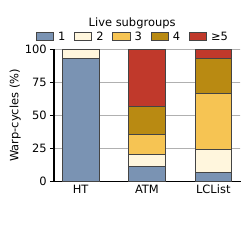}
        \caption{Live-subgroup distribution.}
        \label{fig:case-a1}
    \end{subfigure}
    \hfill
    \begin{subfigure}[b]{0.49\linewidth}
        \centering
        \includegraphics[width=\linewidth]{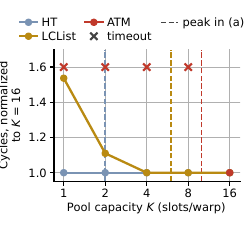}
        \caption{Pool-capacity sensitivity.}
        \label{fig:case-a2}
    \end{subfigure}
    \caption{Subgroup demand and parked-pool capacity.}
    \label{fig:case-a}
\end{figure}

\Cref{fig:case-a}(\subref{fig:case-a1}) shows the fraction of warp-cycles at each live-subgroup
count at $K{=}32$. Demand is highly workload-dependent. With
adjacent lanes sharing buckets, HT has one subgroup in 93.6\% of warp-cycles
(mean $1.06$, peak $2$), while ATM
and LCList spend 42.8\% and 6.5\%, respectively, at five or more.
LCList's hand-over-hand chain holds
two locks per hop, pushing the mean to $3.09$ (peak $6$). ATM's lock-ordered two-account transfers create the deepest simultaneous dependency chains of the three (mean $3.75$, peak $10$), which is still less than a
third of $W$. This supports SIMTurn's central idea that subgroups need to
be tracked only on the workload's divergent paths, not across the full
warp width.

\Cref{fig:case-a}(\subref{fig:case-a2}) plots execution cycles against pool capacity $K$,
normalized per workload to $K{=}16$. A $\times$ marks a 180\,s wall-clock
timeout, and dashed lines mark the measured peaks from
\cref{fig:case-a}(\subref{fig:case-a1}). Capacity $K\ll W$ is enough for
HT and LCList. HT is insensitive down to $K{=}1$, and LCList is within
$1\%$ of unconstrained by $K{=}4$. ATM is different. It never makes
forward progress for $K\le 8$, while $K{=}16$ is cycle-identical to the
unconstrained run. This is exactly where the independently measured peak
of 10 in \cref{fig:case-a}(\subref{fig:case-a1}) predicts the cutoff. A
pool below a workload's true peak concurrency does not just slow SIMTurn
down. It can stall it. Pool size is therefore a real, workload-dependent
parameter, though even ATM's requirement stays a small fraction of $W$.

\paragraph{Case B: Contention scaling.}
We sweep HT's bucket count from $1$ (all 256 work-items on one lock) to
$64$ and hold every other parameter fixed.

\begin{figure*}
    \centering
    \newlength{\caseBheight}\setlength{\caseBheight}{0.24\textwidth}
    \begin{subfigure}[b]{0.312\linewidth}
        \centering
        \includegraphics[height=\caseBheight]{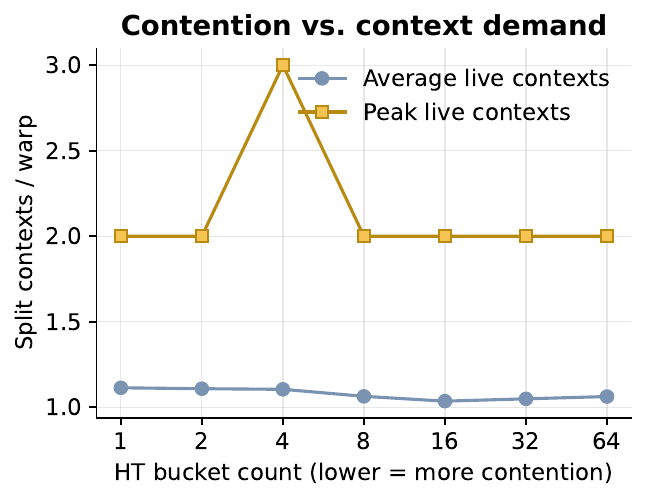}
        \caption{Average and peak live subgroups.}
        \label{fig:case-b1}
    \end{subfigure}\hfill
    \begin{subfigure}[b]{0.314\linewidth}
        \centering
        \includegraphics[height=\caseBheight]{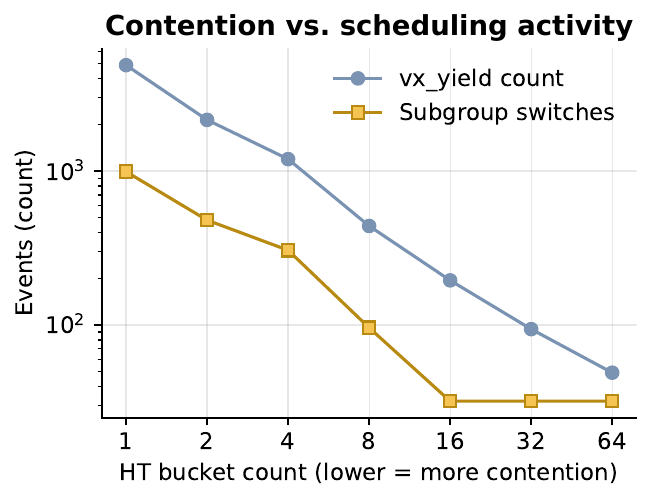}
        \caption{\texttt{vx\_yield} and subgroup-switch counts.}
        \label{fig:case-b2}
    \end{subfigure}\hfill
    \begin{subfigure}[b]{0.344\linewidth}
        \centering
        \includegraphics[height=\caseBheight]{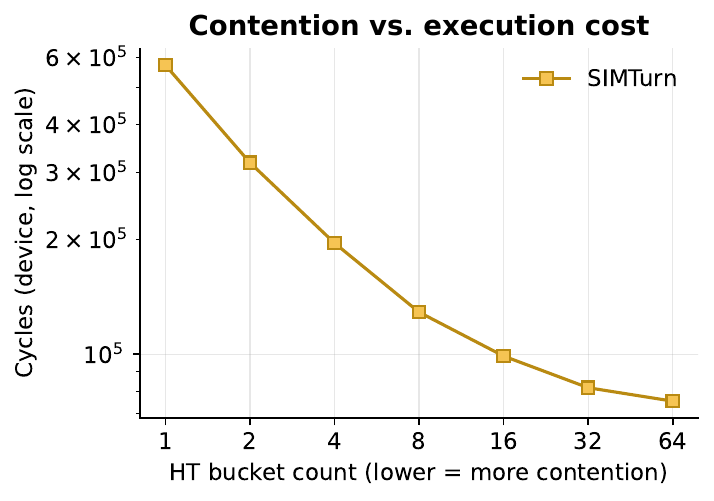}
        \caption{SIMTurn device cycles.}
        \label{fig:case-b3}
    \end{subfigure}
    \caption{HT bucket-count sweep (lower bucket count = more
    contention), fixed $W{=}32$.}
    \label{fig:case-b}
\end{figure*}

\Cref{fig:case-b}(\subref{fig:case-b1}) shows that, for HT, contention barely moves the
\emph{average} number of live subgroups that SIMTurn carries ($1.04$--$1.11$
across the whole sweep). Bucket sharing among adjacent lanes bounds
simultaneous waiters to a small number, roughly independent of contention
even at its maximum. The peak stays at $2$ across the sweep, apart from a
single peak of $3$ at 4 buckets. What contention drives instead is \emph{how often}
the scheduler acts. \Cref{fig:case-b}(\subref{fig:case-b2}) shows the
\texttt{vx\_yield} count falling monotonically from $4878$ at 1 bucket to
$49$ at 64, and subgroup switches falling from $992$ to $32$. These are
$100\times$ and $31\times$ reductions. The floor of $32$ switches comes
from workgroup dispatch and exit, not from contention.
SIMTurn's cost is therefore dominated by scheduling \emph{frequency} under
contention, not by carrying more concurrent state.

\Cref{fig:case-b}(\subref{fig:case-b3}) shows this cost scaling smoothly and monotonically,
from $573$K cycles at 1 bucket down to $75$K at 64. The curve flattens
beyond 16 buckets, where subgroup switches have already reached their
floor. This is a $7.6\times$ range
from contention alone, on the unmodified natural kernel, with every point
correct.

Both case studies point to the same conclusion. Forward progress in this
class of kernel needs a handful of concurrently schedulable subgroups,
almost always far fewer than $W$. Provisioning SIMTurn's pool to that
number, which is measurable and depends on the workload, is sufficient to
recover full performance. The state that \name{} adds scales with how much a kernel
actually diverges, not with per-thread state as in ITS.

\subsection{Hardware Cost}
\label{sec:eval:hardware-cost}

We synthesize the warp scheduler for IPDOM, SIMTurn, and a full per-thread ITS
baseline through Yosys with OpenSTA for timing and power, mapped to the open
ASAP7 (7\,nm, LVT) PDK at an 800\,MHz target clock. We sweep
$W\in\{4,8,16,32\}$ at $N_w{=}8$ and report post-synthesis maximum
frequency, total area (cell $+$ SRAM), and static power for each
configuration. ITS keeps a full PC and mask per lane, while IPDOM and
SIMTurn use the reconvergence stack and the bounded pool, respectively.

\begin{figure}
    \centering
    \begin{subfigure}[b]{0.49\linewidth}
        \centering
        \includegraphics[width=\linewidth]{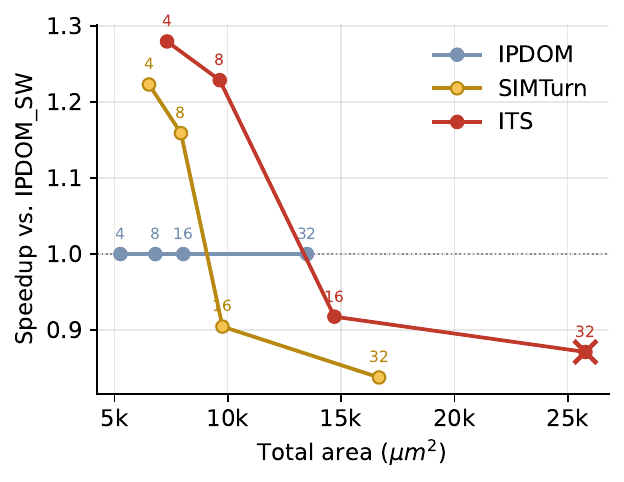}
        \caption{Performance vs. area.}
        \label{fig:asic-pareto}
    \end{subfigure}
    \hfill
    \begin{subfigure}[b]{0.49\linewidth}
        \centering
        \includegraphics[width=\linewidth]{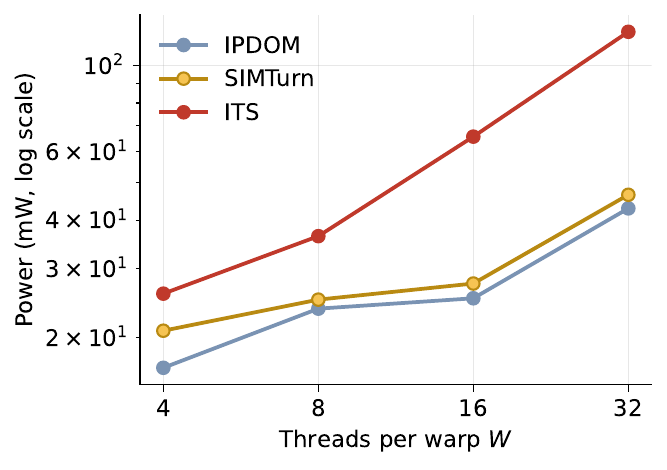}
        \caption{Static power.}
        \label{fig:asic-power}
    \end{subfigure}
    \caption{\name{} ASIC synthesis comparison against IPDOM baseline and ITS.}
    \label{fig:asic-hw}
\end{figure}

\Cref{fig:asic-hw} shows that IPDOM is the cheapest configuration at every
$W$. SIMTurn adds a modest, roughly flat premium of $1.17$--$1.24\times$
IPDOM's area and $1.05$--$1.25\times$ its power. It meets
the 800~MHz target at every width, with a maximum frequency within $7\%$
of IPDOM's. ITS costs far more and grows the fastest,
reaching $1.91\times$ IPDOM's area at $W{=}32$. Its power climbs $4.7\times$ from $W{=}4$
to $W{=}32$, against $2.2\times$ for SIMTurn and $2.6\times$ for IPDOM.
Its maximum frequency falls from $1054$~MHz to $775$~MHz and misses the
target at $W{=}32$.

That extra cost buys nothing past small $W$.
\Cref{fig:asic-hw}(\subref{fig:asic-pareto}) plots the geometric-mean
synchronization speedup against area. At $W{=}4$ and
$W{=}8$, ITS is $1.28\times$ and $1.23\times$ faster than
\texttt{IPDOM\_SW}, and SIMTurn follows within $6\%$ at $1.22\times$ and
$1.16\times$. By $W{=}16$, both fall behind \texttt{IPDOM\_SW}, dragged
down by the single-point-of-contention kernels. Overall, SIMTurn stays close to IPDOM in area,
power, and timing while delivering ITS-level forward progress, without
paying ITS's escalating per-thread cost.

\section{Limitations and Extensions}
\label{sec:discussion}

The evaluated design leaves three extensions open. First, a workload whose
live subgroup count exceeds the pool capacity $K$ stalls, as ATM does for $K\le 8$. Spilling the
oldest parked subgroup to memory, or a scheduling watchdog that falls back
to a safe mode, would remove this limit. Second, a wait that avoids the
forms recognized by \textsc{MayBlock}, such as polling a plain load or calling
through a function pointer, gets
no \texttt{vx\_yield}. Broader compiler analysis or run-time spin
detection in the style of DDOS~\cite{eltantawy2018bows} could cover these
cases. Third, the scheduler uses one fixed policy, pending work first and then
the oldest parked subgroup, and other selection policies are possible.

\section{Conclusion}

We revisit a long-standing tension in SIMT execution where stack-based
reconvergence keeps divergence management cheap but its strict,
depth-first scheduling order can deadlock programs that rely on
intra-warp synchronization. Independent Thread Scheduling resolves this
by giving every thread its own program counter, but at a hardware cost
that scales with warp width rather than with how much a program
actually diverges. We show that a middle ground is possible. \name{}
restores forward progress at the subgroup granularity where divergence
actually occurs, pairing a single new compiler-inserted instruction,
\texttt{vx\_yield}, with a small, bounded, BRAM-backed pool of parked
subgroups in the warp scheduler, and leaving Vortex's existing IPDOM-based reconvergence stack untouched. Across a range of synchronization
microbenchmarks, \name{} reaches ITS-level forward progress on the
evaluated kernels, under the conditions we state, while adding only a modest,
near-flat hardware premium at every evaluated warp width, in contrast to
ITS's cost, which grows steeply with warp width.




\balance
\bibliographystyle{ACM-Reference-Format}
\bibliography{paper}

\end{document}